\documentclass[sigconf]{acmart}

\usepackage{booktabs}
\usepackage{multirow}
\usepackage{amsmath}
\usepackage{enumitem}

\copyrightyear{2026}
\acmYear{2026}
\setcopyright{cc}
\setcctype{by}
\acmConference[AMI '26]{The 1st International Workshop on Agentic Multimodal Intelligence: Models, Benchmarks, and Applications}{November 10--14, 2026}{Rio de Janeiro, Brazil}
\acmBooktitle{The 1st International Workshop on Agentic Multimodal Intelligence: Models, Benchmarks, and Applications (AMI '26), November 10--14, 2026, Rio de Janeiro, Brazil}
\acmDOI{10.1145/3841452.3841492}
\acmISBN{979-8-4007-2941-6/2026/11}

\newcommand{\method}{MM-VeriRec}

\begin{document}

\title{MM-VeriRec: Failure-Guided Fusion for Verifiable Agentic Multimodal Recommendation}

\author{Yufeng Wang}
\affiliation{%
  \institution{Independent Research}
  \country{United States}
}
\email{louiswang524@gmail.com}

\begin{abstract}
Images often carry the recommendation constraints that text metadata only hints at: a movie may need to look dark, a product may need a minimal style, and a visually impossible request should be rejected rather than politely hallucinated. Agentic multimodal recommenders must therefore do more than retrieve plausible items; they must reason over text-image evidence, decide when visual evidence is decisive, and abstain when no valid action exists. We introduce \method{}, a verifiable multimodal recommendation protocol and failure-guided fusion method for hidden visual constraints, image-text mismatch, and impossible-task abstention. \method{} builds tasks from real movie-poster and product-image datasets, verifies each recommendation with deterministic visual attributes, and converts failures into actionable labels such as text-trap following, visual ignorance, and false acceptance. The key idea is that fusion should not merely concatenate modalities; it should diagnose which modality failed and route the next decision through the appropriate repair. Across MM-ML-1M and Amazon Reviews 2023 All\_Beauty, stronger text and vision embeddings improve retrieval but do not remove these failure modes, whereas failure-guided fusion does. On both domains, the adaptive attribute gate reads the same tag family the verifier itself checks; we report its scores (1.0000 on MM-ML-1M, 0.8963 on Amazon All\_Beauty from a 0.5000 plain-fusion baseline) as verifier-aligned upper bounds that test whether the taxonomy routes to the correct repair, not as evidence of open-world visual perception. The more informative evidence is cross-domain transfer under a genuinely non-aligned gate: replacing the oracle tags with an independently derived leave-one-out CLIP detector (per-tag accuracy 0.72 on MM-ML-1M, 0.82 on Amazon) still reaches 0.7028 and 0.6111 visual-grounded success---above both a VBPR multimodal-recommender baseline and plain fusion in both domains---so the failure-guided repair, not the aligned analyzer, drives the gain; the text-versus-visual gap also reproduces across two LLM families, and the repair that helps differs by domain. These results position \method{} as both a benchmark and a practical diagnostic loop for trustworthy agentic multimodal recommendation.
\end{abstract}


\maketitle

\section{Introduction}

Recommendation is becoming visually grounded, but its evaluation is still too often text-shaped. A user asking for a dark-looking thriller, a red floral dress, or a minimalist beauty product is not only asking for a category match; the decisive evidence may live in the item image. Multimodal recommender systems can in principle use this evidence through posters, product photos, and interface previews~\cite{lopezavila2025multimodalrec,abagent2026}. For agentic multimodal intelligence, the question is not only whether a model can perceive images, but whether an agent can act on image evidence reliably, reject visually impossible requests, and explain which part of its multimodal decision process failed.

The hard cases are not merely cases where text is missing; they are cases where text is plausible enough to mislead. Item metadata can claim or imply a visual property that the image does not support, a candidate can satisfy genre or product constraints while missing the visual style, and some requests have no valid answer at all. A reliable agent should therefore do more than rank likely items: it should detect visual contradictions, avoid visually ignorant matches, and abstain under visual impossibility. Aggregate recommendation accuracy blurs these cases together, even though each one asks for a different repair.

\begin{figure*}[t]
\centering
\includegraphics[width=\textwidth]{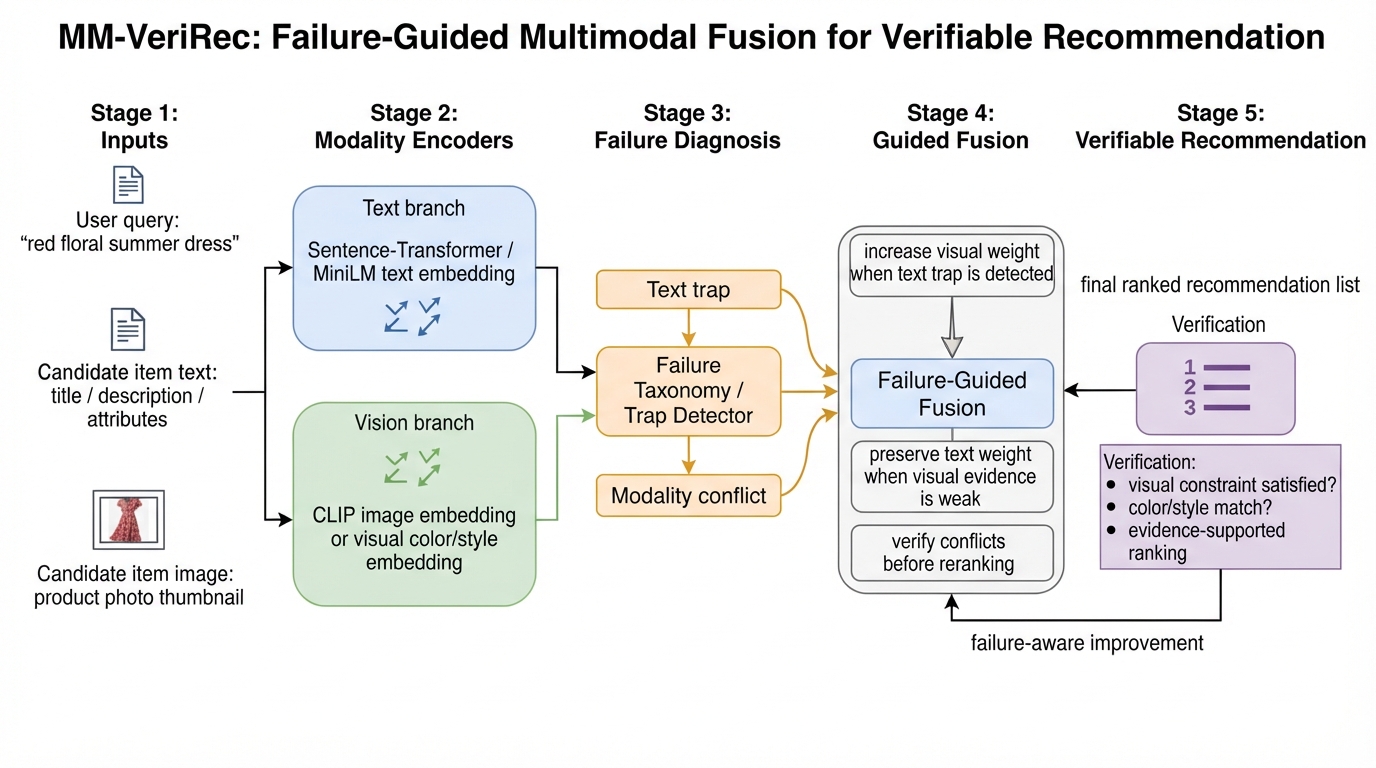}
\caption{Overview of \method{}. The method encodes text and image evidence, diagnoses text traps, visual ambiguity, and modality conflict, then uses those failure labels to guide fusion and verification before producing a ranked recommendation list.}
\Description{A five-stage pipeline diagram for MM-VeriRec. Inputs include a user query, candidate item text, and candidate item image. Text and vision branches produce embeddings. A failure taxonomy detects text traps and modality conflicts. A failure-guided fusion gate adjusts modality weights and feeds a verifiable ranked recommendation list, with feedback from verification to improve fusion.}
\label{fig:method-overview}
\end{figure*}

We introduce \method{}, a benchmark protocol and diagnostic loop for verifiable multimodal recommendation under hidden visual constraints. As summarized in Figure~\ref{fig:method-overview}, \method{} builds tasks from real image-backed recommendation datasets, combines text and visual evidence through cached embeddings or visual attributes, verifies each decision with deterministic visual constraints, and turns the resulting failure labels into fusion decisions. The verifier checks whether a recommendation belongs to the candidate set, satisfies text and rating constraints, avoids consumed items, matches the hidden visual condition, and abstains when no valid candidate exists. This design gives agentic multimodal evaluation a reproducible alternative to open-ended VLM judging while still testing whether image evidence changes the action an agent takes.

The main insight is that multimodal fusion should be failure-guided rather than uniformly multimodal. Simple fusion asks how much text and how much vision to mix. \method{} asks which failure is currently active. False acceptance calls for verification and abstention; text-trap following calls for contradiction penalties; visual ignorance calls for attribute-verified visual gating. This failure-conditioned view makes the benchmark constructive: the evaluation does not stop at saying that an agent failed, but indicates which lightweight repair should be tried next.

The empirical story follows directly from this design. Text-only agents miss hidden visual constraints, generic image fusion can hurt when the visual representation is misaligned with the constraint, and stronger MiniLM/CLIP embeddings still leave residual failures without verification and failure-conditioned repair. These findings support a practical message for efficient representation learning: better embeddings help, but verifiable diagnosis is what turns multimodal evidence into reliable recommendation behavior. In terms of agentic multimodal intelligence, \method{} contributes an \emph{evaluation protocol} for a \emph{trustworthy multimodal agent}: it measures not only whether the agent acts correctly on image evidence, but also whether it abstains under visual impossibility and exposes which cross-modal failure occurred.

Our contributions are:
\begin{enumerate}[leftmargin=*]
    \item \textbf{A verifiable multimodal recommendation protocol} for hidden visual constraints, image-text mismatch, and impossible-task abstention.
    \item \textbf{A failure taxonomy} that separates text-trap following, visual ignorance, false acceptance, and policy failures.
    \item \textbf{A taxonomy-to-intervention fusion controller} that combines cached text-image fusion, visual verification, contradiction penalties, and attribute-verified gating.
    \item \textbf{Cross-domain evidence} on MM-ML-1M movie posters and Amazon Reviews 2023 All\_Beauty product images.
\end{enumerate}

\section{Related Work}
\label{sec:related-work}

\paragraph{Agentic recommender evaluation.}
Agentic recommender benchmarks such as AgentRecBench evaluate LLM-based recommendation agents in interactive scenarios~\cite{shang2025agentrecbench}. tau-Rec moves further toward deterministic verification, hidden intent, tool use, and policy checks~\cite{taurec2026}. \method{} follows this verifiable-agent direction but focuses on a complementary axis: item evidence is multimodal, and correct behavior may require image grounding or abstention under visual impossibility.

\paragraph{Multimodal recommender systems.}
Visual and multimodal recommendation has a long line of work on incorporating image evidence into ranking. VBPR introduced visual factors for implicit-feedback recommendation~\cite{he2016vbpr}, while later neural systems use multimodal graph propagation, latent item structures, bootstrap objectives, and graph denoising to combine visual/textual item features with collaborative signals~\cite{wei2019mmgcn,zhang2021lattice,zhou2023bm3,zhou2023freedom}. These methods primarily optimize ranking quality under observed interactions. \method{} instead asks whether a recommender can satisfy hidden visual constraints, reject impossible requests, and expose which visual-grounding failure occurred.

\paragraph{Agentic and multimodal recommender evaluation.}
Recent surveys and perspective papers describe how LLMs and multimodal signals can support richer recommendation, including semantic reasoning, image evidence, memory, and planning~\cite{huang2025agenticrec,lopezavila2025multimodalrec}. A/B Agent introduces a multimodal movie recommendation sandbox with poster evidence and user-agent simulation~\cite{abagent2026}. Our goal differs from simulation-based A/B testing: \method{} uses multimodal item evidence to create deterministic pass/fail tasks and failure diagnoses.

\paragraph{Pretrained text and vision embeddings.}
General-purpose text and image-text encoders provide strong cached representations for retrieval. Sentence-BERT-style encoders produce semantic sentence embeddings~\cite{reimers2019sbert}, MiniLM compresses transformer representations for efficient inference~\cite{wang2020minilm}, and CLIP aligns images with natural-language supervision~\cite{radford2021clip}. We include MiniLM and CLIP embedding baselines because a benchmark for efficient fusion should test whether stronger representations alone solve the task. Our experiments show that these embeddings improve some baselines but still leave false acceptance, text-image contradiction, and visual ignorance without failure-conditioned repair.

\paragraph{Failure-aware multimodal fusion.}
Agentic multimodal systems often combine cached representations, task-specific evidence, and decision rules before taking an action. This design creates a practical question that standard recommendation metrics do not answer: which visual features are sufficient for a user constraint, and when does fusion fail despite apparently relevant image evidence? \method{} addresses this question by pairing visual-grounded success with failure labels, allowing compact color/style representations, generic CNN embeddings, CLIP image embeddings, and attribute-verified gates to be compared under the same verifier.

\section{Benchmark Design}

\subsection{Task Format}

Each task contains a user history, a candidate catalog, a visible text preference, a hidden visual constraint, and policy constraints. The visible preference covers ordinary recommendation attributes such as genre, product type, and minimum rating. The hidden visual constraint refers to image-derived attributes such as brightness, dominant color, contrast, or visual style density. Some tasks are impossible because no candidate satisfies the visual constraint; in those cases, the correct behavior is abstention.

\subsection{Verifier}

The verifier assigns a decision to exactly one outcome. A recommendation is a valid success only if it is in the candidate set, not already consumed, satisfies the text/rating constraints, and matches the hidden visual attribute. An abstention is valid only when no valid candidate exists. Otherwise, failures are classified into policy violations, text-trap following, visual ignorance, false acceptance, hallucinated items, or over-abstention.

\subsection{Datasets}

We instantiate \method{} on two real image-backed recommendation datasets. MM-ML-1M provides MovieLens-style interactions~\cite{harper2015movielens}, movie metadata, and poster images. We derive visual tags from poster pixels and use user histories to avoid recommending consumed items. Amazon Reviews 2023 All\_Beauty provides user reviews, product metadata, and product images~\cite{amazonreviews2023}. We derive product-image color and style tags, then construct tasks around product terms, ratings, consumed items, and hidden visual constraints.

\subsection{Constraint Generation}

Hidden visual constraints are produced by a deterministic image-analysis pipeline that is fixed before any agent is evaluated. For each item we compute low-level descriptors directly from the poster or product image and discretize them with fixed thresholds into categorical tags: mean brightness (bright or dark), dominant color channel (red, green, or blue), and global color contrast (high or low); for product images we add a style tag (busy or minimal) from per-pixel color variance. A task's hidden constraint is one such tag, and a task is labeled impossible when no catalog candidate carries it after text and rating filtering. Because tags are computed once per item and cached, the verifier checks constraints by table lookup rather than by re-encoding images, which keeps every pass/fail decision deterministic and reproducible.

We deliberately separate two roles that are easy to conflate. The \emph{constraint generator} above defines ground truth and is part of the benchmark. The \emph{agent-side analyzer} used by adaptive guided fusion is a distinct component that an agent may or may not possess. In the main results (Tables~\ref{tab:mm-results} and~\ref{tab:amazon-results}), the adaptive agent's gate reads the constraint generator's own tags on \emph{both} datasets: MM-ML-1M ($1.0000$) and Amazon All\_Beauty ($0.8963$) are therefore both verifier-aligned upper bounds, not claims about open-world perception; the Amazon result is not evidence of a non-aligned analyzer despite the visual representation being cached product-image features. To measure the benchmark's difficulty under a genuinely non-aligned gate, Section~\ref{sec:nonaligned} (Table~\ref{tab:nonaligned}) replaces the oracle tags with an \emph{independently derived} detector: a leave-one-out $k$-nearest-neighbor predictor over cached CLIP image features, computed in a different feature space than the generator's pixel-statistic thresholds and evaluated on each item without ever seeing that item's own label. We report its measured per-tag accuracy ($0.72$ on MM-ML-1M, $0.82$ on Amazon) as the quantified evidence for this independence rather than asserting the detector's errors are uncorrelated with the generator's.

\paragraph{Task and candidate-pool construction.} Each task samples a user with at least 8 (MM-ML-1M) or 2 (Amazon All\_Beauty) prior positive interactions, so that consumed items can be excluded from candidates; a preferred genre or product term and a minimum-rating threshold are drawn from that user's own history to form the visible text preference. The candidate pool for a task is all catalog items matching the visible preference and not already consumed; from this pool we assemble a fixed-size candidate list ($k=18$) mixing items that satisfy the hidden visual constraint, items that carry a misleading text claim about it (traps), and other matching items, so that every task presents both a valid answer (when one exists) and a plausible distractor. Each task is independently marked impossible with probability $0.12$, in which case no candidate in the pool carries the hidden visual tag and the correct action is abstention. Tasks are generated fresh per random seed from the full catalog rather than drawn from a held-out split: MM-ML-1M uses 120 tasks per seed (360 over three seeds) and Amazon All\_Beauty uses 90 tasks per seed (270 over three seeds), with no separate calibration or validation set. This matters for how repair selection should be read: every agent variant described in Section~\ref{sec:methods} (visual verification, contradiction penalty, attribute gate) is a fixed, deterministic scoring rule with no parameters fit to any task's outcome, so there is no train/test leakage to guard against--each seed's tasks are scored once, by a rule chosen before evaluation, not adapted online from observed failures. What we call ``adaptive'' in adaptive guided fusion refers to the taxonomy determining, at design time and once per dataset, \emph{which} fixed repair to deploy (Section~\ref{sec:failure-to-repair} makes this selection explicit); it is not an online controller that reads test-set failure statistics.

\subsection{Evaluation Metrics}

\method{} reports correctness together with failure explanations. Visual-grounded task success measures whether the recommendation satisfies hidden visual intent. Failure counts explain why an agent fails. We also report policy pass rate, abstention accuracy, image-text mismatch pass rate, and visual evidence use when the corresponding task type is present. These metrics keep the benchmark focused on the paper's central question: whether an agentic fusion policy can use image evidence correctly under hidden visual constraints.

\section{Methods}
\label{sec:methods}

\subsection{Agents}

Figure~\ref{fig:method-overview} summarizes the end-to-end protocol. We evaluate LLM-style and embedding-style agents. The DeepSeek text-only agent receives metadata and misleading text visual claims, while the visual-tag agent additionally receives poster-derived visual tags. For embedding retrieval, the lexical baseline uses TF-IDF metadata embeddings, and stronger text baselines use MiniLM and CLIP text embeddings. The plain fusion agent combines text similarity with image similarity. We test generic ResNet-18 poster embeddings, CLIP image embeddings, and aligned color/statistics embeddings.

\subsection{Failure-Guided Fusion}

Failure-guided fusion uses the taxonomy as a controller. The base score for a candidate $i$ is
\[
s_i = \alpha s_i^{text} + (1-\alpha)s_i^{vision}.
\]
The main experiments use a fixed balanced fusion weight $\alpha=0.5$ for all datasets, seeds, and repair variants, rather than tuning $\alpha$ separately for each agent. We similarly keep the visual threshold at $0.50$ and the text-trap penalty at $0.25$ across the main cached-feature experiments. This choice isolates the effect of failure-conditioned repairs; a larger deployment study should tune these parameters on a validation split and report sensitivity curves. Visual verification masks candidates whose visual score falls below the threshold and abstains if no candidate remains. Text-trap suppression subtracts the fixed penalty when metadata claims the requested visual attribute but the image evidence does not support it. Adaptive guided fusion adds an attribute-verified gate when visual ignorance is the dominant residual error:
\[
s_i = -\infty \quad \text{if } i \text{ lacks the required visual attribute.}
\]
This final step uses an aligned visual-attribute analyzer and is therefore a controlled upper-bound style intervention, not a generic claim about arbitrary vision encoders.

\begin{table}[t]
\centering
\caption{Failure-guided fusion controller. Each repair is activated by a measured failure mode rather than applied uniformly.}
\label{tab:controller}
\begin{tabular}{lll}
\toprule
Failure mode & Repair & Cost profile \\
\midrule
False acceptance & Threshold + abstain & score mask \\
Text trap & Penalty & tag check \\
Visual ignorance & Attribute gate & attribute check \\
\bottomrule
\end{tabular}
\end{table}

\subsection{From Failure Labels to Fusion Decisions}
\label{sec:failure-to-repair}

The controller is intentionally simple because the goal is to show how verifiable evaluation can guide efficient fusion design. A monolithic multimodal reranker can hide why a decision improved, and a single fixed fusion weight can overuse image evidence even when the active error is not visual. \method{} instead treats each failure label as a design signal. If false acceptance is frequent, the agent needs a reject option, not a different ranking weight. If text-trap following is frequent, the agent needs contradiction handling, not more text similarity. If visual ignorance is frequent, the agent needs a stronger visual gate before ranking.

\begin{table}[t]
\centering
\caption{Example failure-to-repair logic used by adaptive guided fusion.}
\label{tab:failure-repair}
\begin{tabular}{ll}
\toprule
Observed error & Fusion change \\
\midrule
Impossible recommendation & add abstain gate \\
Misleading metadata & add trap penalty \\
Missed color/style constraint & require attr. match \\
Over-abstention & lower threshold \\
\bottomrule
\end{tabular}
\end{table}

This design also clarifies why the Amazon transfer result matters, and it lets us decompose how much of the final gain is attributable to the taxonomy's contradiction penalty versus the attribute gate. On MM-ML-1M (Table~\ref{tab:mm-results}), adding the text-trap penalty (text+image fusion $\rightarrow$ taxonomy-guided) raises success from 209 to 229 by removing all 44 trap failures, while adding the attribute gate on top (taxonomy-guided $\rightarrow$ adaptive guided) raises success further from 229 to 323 by removing the remaining 94 visual-ignorance failures: the gate accounts for most of the improvement, consistent with the concern that the headline gain is driven by attribute filtering rather than by the taxonomy itself. On Amazon (Table~\ref{tab:amazon-results}), aligned fusion had already eliminated text-trap following, so the penalty step is a no-op (135 $\rightarrow$ 135); the gate step alone (135 $\rightarrow$ 242) removes all 107 visual-ignorance failures and is responsible for the entire adaptive-guided gain. We read the taxonomy's contribution accordingly, not as a source of accuracy by itself, but as the mechanism that correctly identifies, once per dataset before evaluation, which of the two repairs is worth deploying: a contradiction penalty on MM-ML-1M, where text traps are the first residual failure, and only an attribute gate on Amazon, where they are not. This is the central agentic multimodal lesson of the paper: which repair helps is conditional on the failure distribution, even though the repair mechanism (the gate) supplies most of the raw accuracy.

\section{Experiments}

\subsection{Setup and Metrics}

The primary metric is visual-grounded task success. Secondary metrics include policy pass rate, abstention accuracy, image-text mismatch pass rate, visual evidence use, and failure taxonomy counts. All strong claims use three random seeds. We evaluate lexical TF-IDF retrieval, MiniLM sentence embeddings, ResNet-18 image embeddings, CLIP image embeddings, and compact color/statistics features. We use cached image features for embedding and attribute-based agents, so the experiments compare fusion and verification decisions rather than repeated image encoding.

\subsection{MM-ML-1M Main Results}

MM-ML-1M tests whether agents can use poster-derived visual evidence rather than plausible text alone. The DeepSeek text-only agent reaches 0.0444 visual-grounded success, while the visual-tag agent reaches 0.9222, supporting the basic benchmark premise that visual evidence is necessary when the hidden constraint is visual. This gap is not specific to one model family: a second LLM, Gemini~2.5~Flash, reproduces it almost exactly, moving from 0.0222 text-only to 0.9444 with poster-derived visual tags (mean over three seeds). For embedding retrieval, generic image fusion is not automatically beneficial: ResNet-18 poster fusion decreases success relative to text-only retrieval, while aligned color-statistics fusion improves success from 0.3611 to 0.5806. Table~\ref{tab:repr-ablation} gives the representation ablation behind this claim. The compact color-statistics representation is cheaper and better aligned with the benchmark's visual constraints than the generic ResNet-18 embedding, and adaptive guided fusion then adds a lightweight attribute gate on top of aligned features.

\begin{table}[t]
\centering
\caption{MM-ML-1M representation ablation. Compact aligned features outperform generic visual embeddings for verifier-friendly visual constraints. The adaptive-guided 1.0000 is a verifier-aligned upper bound (analyzer aligned with the verifier attributes), not open-world saturation.}
\label{tab:repr-ablation}
\begin{tabular}{llr}
\toprule
Agent & Visual representation & Success \\
\midrule
Text retrieval & none & 0.3611 \\
Text+image fusion & ResNet-18 & 0.1972 \\
Text+image fusion & color statistics & 0.5806 \\
Adaptive guided & color attrs. + gate & 1.0000 \\
\bottomrule
\end{tabular}
\end{table}

Table~\ref{tab:mm-results} shows the main MM-ML-1M fusion results. As a learned multimodal-recommender reference, we train VBPR~\cite{he2016vbpr}, which ranks candidates by BPR preference plus a visual factor over the cached image features. It reaches only $0.4750$ visual-grounded success and still leaves 103 visual-ignorance and 37 false-acceptance failures over 360 tasks: optimizing observed-interaction ranking neither satisfies hidden visual constraints nor triggers abstention on impossible tasks, which is exactly the gap the verifier is designed to expose. Plain fusion improves valid success but leaves text traps, visual ignorance, and false acceptance. Visual verification converts false acceptance into valid abstention. The first taxonomy-guided agent eliminates text-trap following but leaves visual ignorance. Adaptive guided fusion removes the measured visual-ignorance failures by using attribute-verified visual gating. The resulting perfect visual-grounded score should be read as a controlled sanity result: the benchmark attributes and the adaptive gate are deliberately aligned, so this setting tests whether the failure taxonomy can route to the right repair, not whether the task exhausts the difficulty of future multimodal recommenders.

\begin{table}[t]
\centering
\caption{MM-ML-1M failure-guided fusion results over 360 tasks (3 seeds), with per-seed standard deviation on the success rate. Counts sum to 360 in every row once valid abstentions are included. The adaptive-guided perfect success is a controlled verifier-aligned upper bound (the gate reads the constraint generator's own tags); see the non-aligned-detector results (Table~\ref{tab:nonaligned}) for the genuinely non-aligned setting.}
\label{tab:mm-results}
\small
\setlength{\tabcolsep}{3.2pt}
\begin{tabular}{lrrrrrr}
\toprule
Agent & Succ. & Rate (std) & Trap & Ign. & FAcc & Abst \\
\midrule
VBPR~\cite{he2016vbpr} & 171 & 0.475 (0.031) & 49 & 103 & 37 & 0 \\
Text+image fusion & 209 & 0.581 (0.048) & 44 & 70 & 37 & 0 \\
Visual verified & 209 & 0.683 (0.018) & 44 & 70 & 0 & 37 \\
Taxonomy-guided & 229 & 0.739 (0.032) & 0 & 94 & 0 & 37 \\
Adaptive guided & 323 & 1.000 (0.000) & 0 & 0 & 0 & 37 \\
\bottomrule
\end{tabular}
\end{table}

\subsection{Transfer and Stronger Embeddings}

Table~\ref{tab:amazon-results} shows the Amazon All\_Beauty transfer result. The VBPR baseline~\cite{he2016vbpr} again improves over text-only retrieval ($0.4185$ versus $0.3741$) yet leaves 106 visual-ignorance and 28 false-acceptance failures over 270 tasks, confirming that the pattern is not specific to the movie domain. Text-only retrieval reaches 0.3741 visual-grounded success, aligned product-image fusion reaches 0.5000, and visual verification removes false acceptance. The earlier taxonomy-guided penalty is neutral because text traps are already removed by aligned fusion. Adaptive guided fusion targets the active residual failure mode, visual ignorance, and improves to 0.8963; as on MM-ML-1M, this gate reads the constraint generator's own tags, so 0.8963 is a verifier-aligned upper bound rather than a non-aligned transfer result (Table~\ref{tab:nonaligned} reports the genuinely non-aligned CLIP-detector result of 0.6111, still above the 0.5000 plain-fusion baseline). This transfer result matters because the successful repair differs across domains: MM-ML-1M benefits first from contradiction handling, while Amazon primarily needs an attribute gate.

\begin{table}[t]
\centering
\caption{Amazon All\_Beauty visual success and failure counts over 270 tasks (3 seeds), with per-seed standard deviation on the success rate. Counts sum to 270 in every row once valid abstentions are included. The adaptive-guided row is, like MM-ML-1M, a verifier-aligned upper bound: the gate reads the constraint generator's own tags, not an independently derived analyzer (Table~\ref{tab:nonaligned} reports the genuinely non-aligned setting).}
\label{tab:amazon-results}
\small
\setlength{\tabcolsep}{3.2pt}
\begin{tabular}{lrrrrrr}
\toprule
Agent & Succ. & Rate (std) & Trap & Ign. & FAcc & Abst \\
\midrule
VBPR~\cite{he2016vbpr} & 113 & 0.419 (0.032) & 23 & 106 & 28 & 0 \\
Text+image fusion & 135 & 0.500 (0.024) & 0 & 107 & 28 & 0 \\
Visual verified & 135 & 0.500 (0.024) & 0 & 107 & 0 & 28 \\
Taxonomy-guided & 135 & 0.500 (0.024) & 0 & 107 & 0 & 28 \\
Adaptive guided & 242 & 0.896 (0.005) & 0 & 0 & 0 & 28 \\
\bottomrule
\end{tabular}
\end{table}

Table~\ref{tab:strong-embeddings} tests whether the main conclusion is an artifact of weak lexical text retrieval. MiniLM raises text-only success on Amazon from the TF-IDF setting, and CLIP image embeddings improve plain fusion on Amazon. However, stronger embeddings do not remove the structural failures exposed by the verifier. On MM-ML-1M with MiniLM text and CLIP image features, plain fusion still produces 37 false acceptances and 125 visual-ignorance failures over 360 decisions; visual verification removes false acceptance but leaves 124 visual-ignorance failures; taxonomy guidance removes text traps but still leaves 137 visual-ignorance failures. The adaptive gate remains the only tested method that removes visual ignorance in both domains.

\begin{table}[t]
\centering
\caption{Stronger embedding validation. Values are mean visual-grounded success over three seeds, with standard deviation in parentheses.}
\label{tab:strong-embeddings}
\small
\begin{tabular}{llrr}
\toprule
Dataset & Setting & Fusion & Adaptive \\
\midrule
MM & MiniLM + color & 0.564 (0.077) & 1.000 (0.000) \\
MM & MiniLM + CLIP image & 0.392 (0.089) & 1.000 (0.000) \\
Amazon & MiniLM + color & 0.633 (0.045) & 0.896 (0.005) \\
Amazon & MiniLM + CLIP image & 0.656 (0.033) & 0.896 (0.005) \\
\bottomrule
\end{tabular}
\end{table}

\subsection{Non-Aligned Attribute Detector}
\label{sec:nonaligned}

The adaptive gate in the preceding tables reads the verifier's own attribute tags, so its scores on \emph{both} MM-ML-1M and Amazon are verifier-aligned upper bounds, not just MM-ML-1M's. To separate the value of the failure-guided repair from the value of a perfect detector, we replace the oracle gate with two non-aligned tag sources while leaving the verifier unchanged. The first is an \emph{independently derived} detector: a leave-one-out $k$-nearest-neighbor predictor ($k=15$) over cached CLIP image features, a different feature space than the generator's pixel-statistic thresholds, evaluated so that an item's own label is never used to predict its own tags. We do not assume this makes its errors statistically uncorrelated with the generator's; instead we report its measured per-tag accuracy ($0.72$ on MM-ML-1M, $0.82$ on Amazon, well below the oracle's $1.0$) as the concrete, checkable evidence of imperfection. The second corrupts the ground-truth tags per attribute with an independent probability $\epsilon$.

Table~\ref{tab:nonaligned} reports the result. Replacing the oracle with the CLIP detector, whose per-tag accuracy is only $0.72$ on MM-ML-1M and $0.82$ on Amazon, still yields $0.7028$ and $0.6111$ visual-grounded success, well above the plain-fusion baselines of $0.5806$ and $0.5000$. Success degrades smoothly as $\epsilon$ grows, tracking detector accuracy rather than collapsing. This indicates that the taxonomy-guided repair, not the aligned analyzer alone, drives the improvement: the failure-conditioned attribute gate helps even when the detector is imperfect and derived independently of the benchmark's labeling pipeline. The aligned rows reproduce the earlier $1.0000$ and $0.8963$, confirming that the ablation shares the main pipeline and that both are verifier-aligned upper bounds rather than non-aligned transfer evidence.

\begin{table}[t]
\centering
\caption{Non-aligned detector ablation. Adaptive-guided visual-grounded success (mean $\pm$ std over three seeds) as the gate's tag source degrades from the oracle to an independently derived CLIP detector and to $\epsilon$-corrupted tags. The oracle rows are verifier-aligned upper bounds on both datasets, not open-world results; they reproduce the ``Adaptive guided'' rows of Tables~\ref{tab:mm-results} and~\ref{tab:amazon-results} exactly.}
\label{tab:nonaligned}
\begin{tabular}{lrr}
\toprule
Gate tag source & MM-ML-1M & Amazon \\
\midrule
Aligned oracle (upper bound) & 1.000 (0.000) & 0.896 (0.005) \\
Trained CLIP detector & 0.703 (0.022) & 0.611 (0.027) \\
Corrupted, $\epsilon=0.10$ & 0.972 (0.014) & 0.826 (0.050) \\
Corrupted, $\epsilon=0.20$ & 0.906 (0.051) & 0.767 (0.024) \\
Corrupted, $\epsilon=0.30$ & 0.778 (0.042) & 0.770 (0.041) \\
\bottomrule
\end{tabular}
\end{table}

\section{Discussion, Limitations, and Conclusion}

The experiments support a benchmark-and-repair view of agentic multimodal recommendation. Aggregate success shows whether an agent works, but the taxonomy explains why it fails and which repair is appropriate. Stronger MiniLM and CLIP embeddings improve some baselines, yet they do not by themselves solve impossible-task abstention, text-image contradiction, or visual ignorance. Visual verification is useful for impossible tasks but does not solve visual ignorance. Text-trap penalties are useful only when contradiction traps remain active. Attribute-verified gating is powerful when visual ignorance dominates, but its strength depends on the availability and reliability of an aligned visual analyzer. For agentic multimodal systems, this suggests that fusion should be conditional: apply cross-modal checks where the observed failure distribution says they matter.

\method{} also separates offline representation work from online decision work. Offline, the system can compute image embeddings, color/style summaries, and attribute tags once per item. Online, the agent only scores candidates, applies masks, and chooses whether to abstain. This decomposition explains why the negative ResNet-18 result is useful rather than merely a failed baseline: a generic visual embedding can be high capacity and still miss the specific visual condition being tested. In contrast, compact color/style attributes are aligned with the relevant hidden constraints. The representation question is therefore not only how to use stronger multimodal encoders, but also how to select the visual signal that preserves the decision boundary needed by the task.

The benchmark can also be used as a selection tool for future multimodal agents. If a catalog or domain shows mostly false acceptance, then a thresholded verifier may be enough. If it shows mostly text traps, contradiction-aware fusion is the right intervention. If it shows mostly visual ignorance, the system needs either a better visual representation or an explicit attribute detector. This makes \method{} useful before deployment: it diagnoses which module deserves engineering effort.

\method{} is a stress-test benchmark, not a complete recommender benchmark. The visual attributes are deliberately simple and verifier-friendly, such as color, brightness, contrast, and style density; the benchmark is closer to a controlled test of low-level visual-attribute filtering than to open-ended agentic multimodal recommendation, and extending it to richer, compositional, or human-judged attributes is a direct avenue for future work. This simplicity is useful for deterministic diagnosis, but it also means that a perfect MM-ML-1M score does not imply that the benchmark can challenge arbitrary future multimodal architectures. The adaptive agent uses an analyzer aligned with those attributes, so its results should be interpreted as controlled evidence that taxonomy-guided intervention can repair measured failures, not as proof of robust open-world visual perception. Our non-oracle baselines are VBPR and a leave-one-out CLIP detector; we do not evaluate learned modern multimodal recommenders (e.g., MMGCN, LATTICE, BM3, FREEDOM, cited in Section~\ref{sec:related-work}) or dynamic-routing agents as candidates for the recommendation step itself, so the benchmark's practical significance relative to that broader class of methods remains untested and is a natural next experiment. Relatedly, Section~\ref{sec:failure-to-repair}'s decomposition shows that most of the adaptive agent's numerical gain over taxonomy-guided fusion comes from the attribute gate rather than from the taxonomy narrowing which repair to try; the taxonomy's contribution is choosing the right repair per dataset, not supplying additional accuracy on its own, and future work should test whether this diagnose-then-repair pattern still helps when the repair itself is a learned or higher-capacity component rather than a hand-specified gate. Future work should also use human-validated visual labels, richer user preferences, and larger multimodal catalogs.

We presented \method{}, a verifiable multimodal recommendation protocol for hidden visual constraints, image-text mismatch, and impossible-task abstention. The benchmark exposes failures that text-only evaluations miss and decomposes them into actionable failure modes. By mapping those modes to visual verification, contradiction penalties, and attribute-verified visual gating, adaptive failure-guided fusion improves across both movie-poster and product-image settings. The broader lesson is that agentic multimodal evaluation should not only score agents; it should explain failures well enough to repair them. A natural next step is a staged agent that uses the failure-guided controller for clear cases and escalates uncertain cases to a larger multimodal model.

\bibliographystyle{ACM-Reference-Format}
\bibliography{refs}

\end{document}